\documentclass[aps,prl,twocolumn,nofootinbib]{revtex4}
\usepackage{graphicx}% Include figure files
\usepackage{bm}% bold math
\usepackage{amsmath}
\usepackage{amssymb}
\usepackage[colorlinks,pdfstartview=FitH]{hyperref}
\hypersetup{linkcolor=blue,citecolor=blue,filecolor=black,urlcolor=blue}
\newcommand\sect[1]{\pdfbookmark[1]{#1}{bm:#1}\emph{#1.}---}

\newcommand\hp{\bm {\hat p}}

\newcommand\letter{Letter}

\newcommand{\be}{\begin{equation}}
\newcommand{\ee}{\end{equation}}
\newcommand{\bear}{\begin{eqnarray}}
\newcommand{\eear}{\end{eqnarray}}
\newcommand{\ba}{\begin{array}}
\newcommand{\ea}{\end{array}}
\newcommand{\energy}{{\varepsilon}}

\newcommand{\ebeta}{\epsilon'}
\newcommand{\eomega}{\epsilon_\omega}
\newcommand{\ediss}{\epsilon}

\begin{document}

%\preprint{}
\author{Shiyong~Li}
\email{sli72@uic.edu}
\affiliation{Physics Department, University of Illinois at Chicago, Chicago, 
Illinois 60607, USA}
\author{Mikhail A. Stephanov} 
\email{misha@uic.edu}
\affiliation{Physics Department, University of Illinois at Chicago, Chicago, 
Illinois 60607, USA}
\author{Ho-Ung Yee}
\email{hyee@uic.edu}
\affiliation{Physics Department, University of Illinois at Chicago, Chicago, 
Illinois 60607, USA}

\title{Non-dissipative second-order  transport, spin, and pseudo-gauge transformations in hydrodynamics }
\date{November 2020}

\begin{abstract}
  We derive a set of nontrivial relations between second-order transport coefficients which follow from the second law of thermodynamics upon considering a regime close to uniform rotation of the fluid. We demonstrate that extension of hydrodynamics by spin variable is equivalent to modifying conventional hydrodynamics by a set of second-order terms satisfying the relations we derived. We point out that a novel contribution to the heat current orthogonal to vorticity and temperature gradient reminiscent of the thermal Hall effect is constrained by the second law.
\end{abstract}

\maketitle

\sect{Introduction} Relativistic hydrodynamics \cite{Landau:2013fluid}
is an effective description, at large distance and time scales, of
systems in local thermodynamic equilibrium parameterized by slowly
varying profiles of 4-velocity $u^\mu(x)$ ($u_\mu u^\mu =-1$), local
temperature $T(x)$ and chemical potential $\mu(x)$ for a conserved
charge.  The system of equations based on the conservation laws is
closed, and all dynamic information about the system in the
hydrodynamic regime is contained in the hydrodynamic
variables. Relativistic hydrodynamics has been successful in many
branches of physics, in particular, in describing dynamical evolution
of the fireball created in relativistic heavy-ion collisions (RHIC)
\cite{Jeon:2015dfa,Romatschke:2017ejr}.
 
Recent developments include interesting attempts to incorporate spin
polarization of microscopic constituents as an additional hydrodynamic
variable characterizing the system, which led to consideration of
``spin hydrodynamics''
\cite{Florkowski:2017ruc,Hattori:2019lfp,Gallegos:2020otk}. This is
motivated by importance of spin observables in many applications of
hydrodynamics in condensed matter as well as nuclear
physics. Specifically, each event in non-central RHIC collisions carries a
significant amount of initial orbital angular momentum, some of which is transferred to the spin
polarization of observed hadrons
\cite{STAR:2017ckg,Liang:2004ph,Betz:2007kg,Becattini:2007sr,Huang:2011ru,Jiang:2016woz,Sun:2017xhx}.
However, in the strict sense of hydrodynamics, spin polarization of
plasma constituents should also be in local equilibrium, and must be
determined by conventional hydrodynamic variables.~\footnote{Certain
  variants of spin
  hydrodynamics \cite{Hattori:2019lfp} could
  describe off-equilibrium dynamics of spin polarization in a system
  where relaxation time of spin polarization is much slower than other
  microscopic time scales. Similar extensions of hydrodynamics by
  non-hydrodynamic, but nevertheless {\em parametrically} slow, variables
  have been termed Hydro+ \cite{Stephanov:2017ghc}.}

In this work we assume the standard local equilibrium, and show that
the spin hydrodynamics and the conventional hydrodynamics are two
equivalent descriptions of the same system.  This not only reconciles
the two formulations, but also leads us to find new
constraints for certain transport coefficients in conventional
second-order hydrodynamics.

The central question we answer in this work is the meaning of
pseudo-gauge transformations
\cite{Becattini:2012pp,Becattini:2018duy,Florkowski:2018fap,Speranza:2020ilk} in spin
hydrodynamics.  Since hydrodynamics is based on local thermodynamics,
this question can only be answered after properly addressing how
thermodynamics transform under pseudo-gauge transformations. We show
the equivalence of local thermodynamics between the spin and
conventional hydrodynamics, which requires us to generalize
pseudo-gauge transformation to currents of entropy and conserved
charge. We use these results to prove the equivalence between the spin
hydrodynamics and the conventional hydrodynamics. In particular, we
find that the ideal limit of spin hydrodynamics is equivalent to the
conventional hydrodynamics with certain non-dissipative second-order
transport coefficients. Moreover, five of these
second-order transport
coefficients are uniquely determined by two thermodynamic functions, one
of which appears as the spin
susceptibility in the spin hydrodynamics description.

The existence of such constraints on certain second-order transport
coefficients is an interesting fact by itself, independent of its physics
connection to the spin hydrodynamics. Within the conventional
hydrodynamics, we show that the same constraints can be derived
directly using the second law of thermodynamics, and are therefore
universal. Our derivation is based on a new power counting scheme for
gradients of hydrodynamic variables, motivated by considering small
deviations from one of the equilibrium states of uniformly rotating
fluid, which exist due to conservation of total angular momentum.

We consider {\em dissipative\/} gradients of fluid velocity and of
$\alpha=\mu/T$, as being much smaller than the vorticity and the
temperature gradients neither of which appear in the entropy
production rate at leading order in gradients. This allows us to
reorganize the naive gradient expansion in the entropy production rate
and to derive a set of nontrivial constraints on certain
second-order transport coefficients by applying the second law of
thermodynamics. Our method should be more generally applicable to some
higher-order transport coefficients, as well as to transport
coefficients involving external electromagnetic fields, but we leave
such generalizations to future work.

Although similar constraints have been found for chargless fluid
\cite{Bhattacharyya:2012nq,Jensen:2012jh,Banerjee:2012iz} and charged
fluid in Ref.\cite{Bhattacharyya:2014bha} using different approach,
the constraints in Ref.~\cite{Bhattacharyya:2014bha} appear to be less
stringent, leaving four unconstrained parameters in constrast to the
two coefficients we find. It would be interesting to establish
relationship between the constraints we derive and the ones in
Ref.~\cite{Bhattacharyya:2014bha}, which appears to be a nontrival
task due to difference in choices of variables and frames (we use
conventional Landau frame).

\sect{Non-dissipative second-order hydrodynamics}\hskip 3em\mbox{}
Guided by the observation that vorticity in a uniformly rotating fluid can take
arbitrary values without entropy production, we consider fluid states
 where vorticity and temperature gradients,
$\omega_{\mu\nu}={1\over 2}(\partial^\perp_\mu
u_\nu-\partial^\perp_\nu u_\mu)$, $\partial^\perp_\mu\beta$, while still
being small, are larger than other, dissipative gradients,
$\theta_{\mu\nu}={1\over 2}(\partial^\perp_\mu
u_\nu+\partial^\perp_\nu u_\mu)$ and $\partial^\perp_\mu\alpha$, where
$\partial^\perp_{\mu}\equiv \Delta_{\mu\nu}\partial^\nu$ with
$\Delta_{\mu\nu}=u_\mu u_\nu+g_{\mu\nu}$, $\beta\equiv 1/T$ and $\alpha\equiv \beta\mu$.  To this end, we introduce
the following power counting scheme: $\omega_{\mu\nu} \sim \eomega$,
$\partial_\mu^\perp \beta \sim \ebeta$,
$\theta_{\mu\nu}\sim\partial^\perp_\mu\alpha\sim\ediss$, while any
further spatial derivative on $(\omega_{\mu\nu},\beta)$ and
$(\theta_{\mu\nu},\alpha)$ brings an extra $\ebeta$ and $\ediss$,
respectively. For example,
$\partial^\perp_\rho \omega_{\mu\nu} \sim \eomega\ebeta$,
$\partial^\perp_\mu\partial^\perp_\nu \beta \sim \ebeta^2$, and
$\partial^\perp_\rho \theta_{\mu\nu} \sim
\partial^\perp_\mu\partial^\perp_\nu \alpha\sim \ediss^2$. In
addition, we consider spatial gradients of thermal vorticity to be of the
same order as the dissipative gradients,
i.e. $\partial^\perp_\nu(\beta\omega^\mu)\sim \eomega\ediss$ rather
than $\eomega\ebeta$, which means
$\partial_\nu^\perp \omega^\mu = -(\partial^\perp_\nu \beta)\omega^\mu
/\beta + {\cal O}(\eomega\ediss)$.
From this and the ideal equation
of motion, one can show that
$\partial_\mu\omega^\mu\sim \omega^\mu\partial_\mu\beta\sim
\eomega\ediss$.

We then invoke the hierarchy,
$\ebeta^2 \ll \ediss \ll \eomega\ebeta \ll \eomega^2 \ll \ebeta \ll
\eomega \ll 1$.
As we will see, this allows us to focus on the
vorticity related terms arising from certain second-order transport
coefficients as the leading contributions to the entropy production
rate up to order $\eomega\ebeta \ediss$, while the dissipative terms
from first-order transport coefficients are of order
$\ediss^2\ll \eomega\ebeta \ediss$, and are thus sub-leading.  Note that
$\eomega\ebeta \ediss$ would naively be of higher order than
$\ediss^2$ in the conventional gradient expansion.  By careful
inspection of all possible terms in the entropy production rate,
potentially larger terms of $\eomega^4$, $\eomega^3\ebeta$ and
$\eomega^2\ebeta^2$ can be shown to be absent in  parity even plasma
that we focus on in this work. Then the second law of thermodynamics,
i.e. the non-negativity of entropy production, should be applied to
these leading contributions involving second-order transport
coefficients.

We write the general parity even constitutive relations for symmetric
energy-momentum tensor, as well as for charge and entropy currents:
\bear
&T^{\mu\nu} = (\energy + p)u^\mu u^\nu + pg^{\mu\nu} +\Delta T^{\mu\nu}\label{NonDissipativeSecondOrderEMTensor}, 
\\
&j^\mu = nu^\mu + \Delta j^\mu, \\
&s^\mu = su^\mu + \Delta s^\mu,
\eear
where $\Delta T^{\mu\nu}$, $\Delta j^\mu$ and $\Delta s^\mu$ contain
all relevant second order terms in our hierarchy,
\begin{align}
    \Delta T^{\mu\nu} &=  a_0\Delta^{\mu\nu}\omega^{\lambda\rho} \omega_{\lambda\rho} + a_1 \omega^\mu_{\phantom\mu\lambda}\omega^{\lambda\nu},\label{SecondOrderTensor}  \\
    \Delta j^\mu &= c_1\Delta^\mu_\rho\partial_\nu\omega^{\nu\rho} + c_2\omega^{\mu\nu}\partial_\nu \beta, \label{SecondOrderCurrent}  \\
    \Delta s^\mu + \alpha\Delta j^\mu &=
    b_1\Delta^\mu_\rho\partial_\nu\omega^{\nu\rho} +
    b_2\omega^{\mu\nu}\partial_\nu \beta +
                                        b_3\omega^{\mu\nu}\partial_\nu
                                        \alpha,
                                        \label{SecondOrderEntropy}
\end{align}
with seven second-order transport coefficients $\{a_i, b_i, c_i\}$. We
do not need to include the first-order transport terms as explained
above, and we omit other possible second-order terms, such as
$\partial^\perp_\mu\beta\partial^\perp_\nu \beta$ in
$\Delta T^{\mu\nu}$ and $\omega^{\mu\nu}\partial_\nu\alpha$ in
$\Delta j^\mu$, that do not contribute to the entropy production rate
to order $\eomega\ebeta\ediss$, and whose coefficients are thus
not constrained by our method.  We also remark that one could put a
purely spatial gradient
$\Delta^\mu_\rho\Delta^\gamma_\nu\partial_\gamma \omega^{\nu\rho}$ in
place of $\Delta^\mu_\rho\partial_\nu\omega^{\nu\rho}$, but this would
be equivalent up to a redefinition of $\{b_2,c_2\}$ due to the ideal
equations of motion and the thermodynamic relation
${\beta }dp = - w d\beta +{n}d\alpha$.

Introducing $\omega^\mu\equiv{1\over
  2}\ediss^{\mu\nu\alpha\beta}u_\nu\omega_{\alpha\beta}$ and using
the identity
 \begin{equation}
\omega_\mu D\omega^\mu =
{\omega^{\mu\nu}\frac{(\partial_\mu \energy)(\partial_\nu p)}{2w^2}}
-{\omega_\mu\omega^\mu \frac{D    p}{w}}
-\omega_\alpha^{\,\,\mu}\omega^{\alpha\nu}\theta_{\mu\nu}\label{eq:oDo}
\end{equation}\\
 which follows from the ideal equations of motion, where $w=\energy+p$
 and $D\equiv u\cdot\partial$, one finds the entropy production rate
 up to $O(\eomega\ebeta\ediss)$ given by
\begin{widetext}
\begin{equation}
\partial_\mu s^\mu = C^{(1)}\omega_\nu\omega^\nu\theta + C^{(2)}(\partial^\perp_\nu\omega^{\nu\mu})\partial^\perp_\mu\alpha +C^{(3)}(\partial^\perp_\nu\omega^{\nu\mu})\partial^\perp_\mu\beta + C^{(4)}(\partial_\mu\beta)\omega^{\mu\nu} (\partial_\nu\alpha)  + C^{(5)}\theta_{\mu\nu}\omega_{\alpha}^{\,\,\,\mu}\omega^{\alpha\nu} 
, \label{entropyProductionForRatatingSysterm}
\end{equation}
\end{widetext}
where $\theta\equiv\theta^\mu_\mu=\partial\cdot u$ and $C^{(i)}$ are given by
\begin{subequations}
  \begin{align}
  C^{(1)} &= -2(a_0\beta + b_1 + 2b_1c_s^2 + b_2 w \beta_\energy + b_3\alpha_p w c_s^2), \label{C1} \\
  C^{(2)} &= \left({\partial b_1 \over \partial \alpha}\right)_{\!\!\beta}
  + b_3 -c_1,\quad
  C^{(3)}=\left({\partial b_1 \over \partial \beta}\right)_{\!\!\alpha} + b_2,\label{C3} \\
  C^{(4)} &=   {b_3\over \beta} + \left({\partial b_3 \over \partial \beta}\right)_{\!\!\alpha} + {n\over \beta w}\left({\partial b_1 \over \partial \beta}\right)_{\!\!\alpha} + {1\over \beta}\left({\partial b_1 \over \partial \alpha}\right)_{\!\!\beta} \nonumber \\
  &+ 2b_1 {\partial \over \partial \beta}\left({n\over \beta w}\right)_{\!\!\alpha}  + {b_2 n\over \beta w}-  \left({\partial b_2 \over \partial \alpha}\right)_{\!\!\beta}  -{c_1\over \beta} + c_2, \label{C4} \\
  C^{(5)} &= a_1 \beta + 4b_1, \label{C5}
  \end{align}
\end{subequations}
where $c_s^2 =\left({ \partial p / \partial
    \energy}\right)_{s/n}$, $\alpha_p = \left({\partial \alpha/
    \partial p}\right)_{s/n}$ and $\beta_\energy =\left({\partial
    \beta / \partial \energy}\right)_{s/n}$ are thermodynamic
derivatives taken with ${s / n}$ fixed, which appear naturally due to the ideal equations of motion, $(u\cdot\partial) (s/n) =0$. 

All five terms in Eq.~(\ref{entropyProductionForRatatingSysterm}) are independent and can have either sign for generic initial conditions.
The second law of thermodynamics thus requires that all $C^{(i)}$ vanish.
This gives five constraints for seven unknowns $\{a_i, b_i, c_i\}$,
which determines them up to two free functions. Choosing $a_0$ and
$a_1$ as two given functions, one can solve for the other five
transport coefficients  without any integration, proceeding in the
following order: 
\begin{subequations}
\label{bia1}
  \begin{gather}
    b_1 = - {\beta a_1\over 4}, \label{eq:solutionb1b2}\quad
    b_2 = -\left(\partial b_1 \over\partial
      \beta\right)_{\!\!\alpha},\,\\
    b_3={1\over \alpha_p w c_s^2}\left[
      w\beta_\epsilon
      \left({\partial b_1 \over \partial\beta}\right)_{\!\!\alpha}
      - b_1- 2b_1c_s^2 -\beta a_0 \right], \label{solutionForb_3}\\
    c_1 = b_3 + \left({\partial b_1 \over \partial \alpha}\right)_{\!\!\beta},
    \quad
    c_2 = -\left({\partial c_1 \over \partial \beta}\right)_{\!\!\alpha}
    - 2b_1 {\partial \over \partial \beta }\left({n \over \beta w}\right)_{\!\!\alpha}. \label{solutionForc_1andc_2}
  \end{gather}
\end{subequations}

\pdfbookmark[2]{Conformal field theory}{bm:cft}

As a nontrivial check of these relations we can consider conformal
theory, such as the strongly coupled conformal plasma described by
AdS/CFT correspondence for which some of the coefficients have been
calculated in Ref.~\cite{Erdmenger:2008rm}.  Conformal invariance
imposes certain constraints on some of the thermodynamic quantities,
such as $w=4\energy/3$, $c_s^2=1/3$,
$\beta_\energy= -\beta/(4\energy)$, $\alpha_p=0$, as well on transport
coefficients: $a_1=3a_0$ and $\left({\partial b_1/
    \partial\beta}\right)_\alpha = -b_1/\beta$.   Substituting into
Eq.~(\ref{solutionForb_3}) we find that it is satisfied for any~$b_3$
because, while $\alpha_p=0$, also the expression in the square brackets
nontrivially vanishes, provided~$b_1$ is given by
Eq.~(\ref{eq:solutionb1b2}). Furthermore, conformal invariance
requires
$({\partial}\left({n / \beta w}\right)/ \partial
\beta)_\alpha=0$. Substituting into Eq.~(\ref{solutionForc_1andc_2}),
we find a relationship between $c_1$ and $c_2$ which coincides with a
nontrivial constraint imposed by conformal Weyl symmetry
\cite{Baier:2007ix,Erdmenger:2008rm}. Finally, solving Eqs.~(\ref{eq:solutionb1b2})
and~(\ref{solutionForc_1andc_2}) we can now predict the values of
$b_1$, $b_2$ and $b_3$ which have not been calculated in
Ref.~\cite{Erdmenger:2008rm}, in terms of $a_1$ and $c_1$ which have
been calculated.

\sect{Spin hydrodynamics} Spin hydrodynamics is based on the 
energy-momentum tensor $\Theta^{\mu\nu}$ and the rank-3  tensor
$S^{\mu\alpha\beta}=-S^{\mu\beta\alpha}$ of spin current. The total angular momentum tensor consists of the
orbital and the spin parts,
$J^{\mu\alpha\beta}=(x^{\alpha}\Theta^{\mu\beta}-x^{\beta}\Theta^{\mu\alpha})+S^{\mu\alpha\beta}$,
and the formalism needs the additional conservation law,
$\partial_\mu J^{\mu\alpha\beta}=0$, corresponding to the introduction
of additional spin degrees of freedom. This relates the anti-symmetric part of
$\Theta^{\mu\nu}$ to non-conservation of spin due to spin-orbit exchange
of angular momentum:
$\Theta^{\mu\nu}-\Theta^{\nu\mu}=-
\partial_{\alpha} S^{\alpha\mu\nu}$.

The constitutive relations
are given by
\be
%\begin{align} \label{canEnergy_MomentumWithSpin}
\Theta^{\mu\nu} = \energy u^{\mu}u^{\nu}+p\Delta^{\mu\nu}+(u^{\mu}q^{\nu}+u^{\nu}q^{\mu})+\tau^{\mu\nu} -\frac{1}{2}\partial_{\alpha}S^{\alpha\mu\nu},
\ee
\be
j^\mu = n u^\mu+\tau^\mu, \quad
S^{\mu \alpha\beta} = u^{\mu}S^{\alpha\beta} + \sigma^{\mu\alpha\beta},
%\end{align}
\ee where we do not assume that $u^\mu$ is the Landau frame, $q^{\mu}$
($u\cdot q=0$) is a contribution to energy current, $S^{\mu\nu}$ is
the spin density in local rest frame satisfying the Frenkel condition
$u_{\mu}S^{\mu\nu}=0$, and
$(\tau^{\mu\nu},\tau^\mu,\sigma^{\mu\alpha\beta})$ are dissipative
gradient corrections.  We will not be concerned with these dissipative terms in
our subsequent discussion of an ideal limit, because their inclusion
will not affect our main conclusion.

Writing the entropy current as $s^{\mu}=su^{\mu}+\Delta s^{\mu}$,
($u_\mu\Delta s^\mu=0$) and adding $0=\beta_\nu \partial_\mu
\Theta^{\mu\nu} +\alpha\partial_\mu j^\mu$ to $\partial_\mu s^\mu$ we
obtain the following expression for the entropy production rate:
\begin{widetext}
\bear
\partial_\mu s^\mu   =[Ds -\beta D \energy+\alpha Dn + \frac{1}{2}\beta \omega_{\mu\nu} D S^{\mu\nu} ] + \theta[ s-\beta (\energy+ p) +\alpha n+ \frac{1}{2}\beta \omega_{\mu\nu}S^{\mu\nu} ] -\beta\tau^{\mu\nu}\theta_{\mu\nu}-\tau^\mu\partial_\mu\alpha\nonumber \\
+ \partial_{\mu}[\Delta s^{\mu} -\frac{1}{2}\beta_{\nu}\partial_{\alpha}S^{\alpha\mu\nu} -\beta q^{\mu}+\alpha\tau^\mu+\frac{1}{2}(\partial_{\rho}\beta_{\delta})\sigma^{\mu\rho\delta}] 
+ [(-\beta Du_{\nu} + \partial_{\nu}\beta)(q^{\nu} -\frac{1}{2\beta}S^{\nu\rho}\partial_{\rho}\beta)]- \frac{1}{2}(\partial_{\alpha}\partial_{\mu}\beta_{\nu})\sigma^{\alpha\mu\nu},\label{entropyProductionForSpinHydro}
\eear
\end{widetext}
where $\beta_\nu \equiv \beta u_\nu$.

There exists an ideal limit of spin hydrodynamics
where the right hand side of Eq.(\ref{entropyProductionForSpinHydro}) vanishes. The vanishing of the first two square brackets leads to the following thermodynamics relations \cite{Becattini:2009wh}, 
\be
ds = \beta d \energy -\alpha dn - \frac{\beta}{2} \gamma_{\mu\nu} dS^{\mu\nu},\,
s= \beta(\energy +p)-\alpha n -{\beta\over 2}\gamma_{\mu\nu}S^{\mu\nu}, \label{ThermodynamicRelationOfSpinHydro}
\ee
where the entropy density is a function of $\energy$, $n$ and $S^{\mu\nu}$, with the spin potential being equal to the fluid vorticity in local equilibrium: $\gamma_{\mu\nu}=\omega_{\mu\nu}$. We emphasize that the spin density should be fixed by the spin potential as a thermodynamic relation in equilibrium, i.e. $S^{\mu\nu}=\chi\gamma^{\mu\nu}$ with the spin susceptibility $\chi$ \cite{Aristova:2016wxe}. This determines the spin density in terms of hydrodynamic variables, $S^{\mu\nu}=\chi\omega^{\mu\nu}$.

Vanishing of other terms requires \be
\Delta s^{\mu} = \frac{1}{2}\beta_{\nu}\partial_{\alpha}(u^\alpha S^{\mu\nu}) + \beta q^{\mu}-\alpha\tau^\mu-\frac{1}{2}(\partial_{\rho}\beta_{\delta})\sigma^{\mu\rho\delta}, \label{SpinHydroEntropycurrent}
\ee
and the following relation
\begin{equation}
 q^{\mu}-{w\over n}\tau^\mu = {1\over
   2\beta}S^{\mu\nu}\partial_{\nu}\beta={\chi\over
   2\beta}\omega^{\mu\nu}\partial_\nu\beta. \label{heatCurrent}
\end{equation}

Eq.(\ref{heatCurrent}) is independent of the choice of the
hydrodynamic frame $u^\mu$. However, one can show, by introducing an
impurity~\cite{Rajagopal:2015roa}, that $\tau^\mu$ vanishes in the ``no-drag frame"~\cite{Stephanov:2015roa}.
This is a non-trivial example, similar to Chiral Vortical Effect \cite{Rajagopal:2015roa,Stephanov:2015roa}, where the entropy flows past a static impurity without generating a drag.
One could refer to this non-dissipative heat current we find as the vorticity driven thermal Hall effect. 

\pdfbookmark[2]{Chiral Kinetic Theory}{ckt}

As a nontrivial check of Eq.~(\ref{heatCurrent}) we can calculate the
heat current in the no-drag frame for the microscopic chiral kinetic theory of
massless Dirac fermion. As detailed in Ref.~\cite{Chen:2015gta}, we
choose the fluid rest frame as the spin frame $n_\mu=u_\mu$ so that
the Frenkel condition is satisfied.  With $n^\mu=(1,0,0,0)$, the spin
density $\bm s$ is proportional to the axial current,
$ s^i=\hbar j^i_5=\hbar\bar\psi\gamma^i\gamma_5\psi$. Therefore,
$\bm s =\int_{\bm p,\lambda} \hbar\lambda \bm j_p$, where $\bm j_p$ is
the phase space (Liouville) current and
$\int_{\bm p,\lambda}\equiv\sum_{\lambda=\pm1/2}\int {d^3\bm
  p}/{(2\pi\hbar)^3}$ includes the sum over helicities
$\lambda$. According to Ref.~\cite{Chen:2015gta}, to order
$\mathcal O(\hbar)$,
$\bm j_p = \left(\hat{\bm p} - ({\hbar\lambda}/{p_0})\hp \times \bm
  \nabla\right) f_{\rm eq}$, where $p_0=|\bm p|$.  The second term in
$\bm j_p$ not only accounts for ${2}/{3}$ of the Chiral Vortical
Effect \cite{Chen:2014cla}, but also plays an important role below to
give the correct spin density.  For uniformly rotating (shear-free)
fluid in thermodynamic equilibrium, the particle distribution function
in the no-drag frame takes the form
\cite{Chen:2015gta,Stephanov:2015roa},
$f_{\rm eq}=1/(\exp\{\beta (-p\cdot u +
({1}/{2})S_n^{\mu\nu}\omega_{\mu\nu})\}+1)$, where
$S_n^{\mu\nu}= \lambda \epsilon^{\mu\nu\alpha\beta} p_\alpha
n_\beta/(p \cdot n) $ and $\mu=0$ for simplicity. The spin density
$S^{ij}$ can then be computed as
    \begin{equation}
      S^{ij}= \epsilon^{ijk} s_k =\frac{\omega^{ij}}{24\hbar\beta^2}  +O(\hbar^0)\,.\label{eq:Sij}
    \end{equation}

    On the other hand, the energy-momentum tensor is given by
    $\Theta^{\mu\nu}=\int_{\bm p,\lambda} j_{p}^{\mu}p^{\nu}$.  Using the
    known result for $j_p^{\mu}$, now up to $O(\hbar^2)$ from
    Ref.~\cite{Gao:2018wmr},
\begin{multline}
  \bm j_p = \Bigg( \hp
    - \frac{\hbar\lambda }{p_0}\hp \times \bm  \nabla
    +  \frac{(\hbar\lambda)^2}{p_0^2}  (\hp \times \bm
    \nabla) \times \bm \nabla\Bigg) f_{\rm eq}
  \\
    + {(\hbar\lambda)^2}\bm p \left\{ \bm p \cdot \left[ (\hp \times \bm
    \nabla) \times \bm \nabla\right] \right\}
    \frac{\partial}{\partial p_0}
    \left(\frac{f_{\rm eq}}{2p_0^3}\right),
  \end{multline}
and  $j_p^0 = \hp\cdot\bm j_p$, we find that the symmetric part of
$\Theta^{0i}$ contains the vorticity driven thermal Hall effect
\begin{equation}
  q^{i}
  =\frac{1}{2}\int_{\bm p,\lambda}(j_p^0p^i + j_p^i p^0)
  = \frac{\omega^{ij}\partial_j \beta}{48\hbar\beta^3}
  + \mathcal O(\hbar^0).
\end{equation}
Combined with Eq.~(\ref{eq:Sij}), this agrees  with Eq.~(\ref{heatCurrent}). 
It can also be checked that a similar term in the charge current $\bm\tau=\int_{\bm p,\lambda} \bm j_p$ vanishes, in accordance with our expectation in the no-drag frame.

\sect{Equivalence between spin hydrodynamics and non-dissipative
  second-order hydrodynamics}
It is well known that the energy-momentum tensor in spin hydrodynamics can be
transformed into the symmetric Belinfante-Rosenfeld energy-momentum
tensor by a pseudo-gauge transformation with
$\Sigma^{\alpha\mu\nu}=S^{\alpha\mu\nu}$
\cite{Becattini:2012pp,Becattini:2018duy,Florkowski:2018fap,Speranza:2020ilk}, \bear
\tilde T^{\mu\nu} &=& \Theta^{\mu\nu} + {1\over 2}\partial_\alpha\left(\Sigma^{\alpha\mu\nu} - \Sigma^{\mu\alpha\nu} - \Sigma^{\nu\alpha\mu}\right)\nonumber\\
&=&\frac12\left(\Theta^{\mu\nu}+\Theta^{\nu\mu}\right)-{1\over
  2}\partial_\alpha\left(S^{\mu\alpha\nu}
  +S^{\nu\alpha\mu}\right)\label{BF}.  \eear As a result, the spin
tensor no longer appears in the total angular momentum tensor, i.e.,
$\tilde S^{\alpha\mu\nu} = S^{\alpha\mu\nu} -
\Sigma^{\alpha\mu\nu}=0$. This leaves the conservation of energy and
momentum unchanged,
$\partial_\mu\tilde T^{\mu\nu}=\partial_\mu \Theta^{\mu\nu}=0$, and the
two descriptions of the system based on each energy-momentum tensor
should be equivalent. This suggests that the corresponding
hydrodynamic descriptions based on the same premise of local
equilibrium, i.e. the spin hydrodynamics and the conventional
hydrodynamics, should also be equivalent to each other. We will
establish this equivalence and show that the hydrodynamic variables
between the two descriptions are related quite non-trivially. In the
following, quantities in the spin hydrodynamics will be denoted
without tilde symbol, while those in the conventional hydrodynamics
will be written with tilde symbol.

A central question in showing the equivalence is how the first law of
thermodynamics used in hydrodynamics transforms under the pseudo-gauge
transformation.  The observation crucial for answering this question
is that we can generalize the pseudo-gauge transformation to the
currents of charge and entropy, without affecting their
conservation
\begin{subequations}\label{eq:js-transform}
  \bear
  \tilde j^\mu&=&j^\mu-\partial_\nu\left({a\over 2\chi}S^{\mu\nu}\right)=j^\mu-{1\over 2}\partial_\nu(a\omega^{\mu\nu}), \\
  \tilde s^\mu&=&s^\mu-\partial_\nu\left({b\over
    2\chi}S^{\mu\nu}\right)=s^\mu-{1\over 2}\partial_\nu(b\omega^{\mu\nu}),
  \eear
\end{subequations}
with thermodynamic functions $a(\energy,n)$, $b(\energy,n)$.  An
intuitive understanding of physics of these transformations is
obtained by noting that the spatial part of
$-\partial_\nu(a\omega^{\mu\nu} )/2$ can be interpreted as the
magnetization current $\bm\nabla\times\bm M$ with  vorticity induced
magnetization $\bm M=-a\bm\omega/2$, i.e. the Barnett effect.

Since the local charge and entropy densities, $(n,s)$, are defined by
$n=-u_\mu j^\mu$ and $s=-u_\mu s^\mu$ respectively, transformations in
Eqs.~(\ref{eq:js-transform}) redefine them 
$\tilde n=n-\Delta n$,
$\tilde s=s-\Delta s$,
where
\begin{equation}
\Delta n = -{1\over
  2}u_\mu\partial_\nu(a\omega^{\mu\nu})
%=-{1\over 2}a\omega_{\mu\nu}\omega^{\mu\nu}
=-a\omega_\mu\omega^\mu, \quad
\Delta s = -b\omega_\mu\omega^\mu.\label{eq:DnDs}
\end{equation}
Taking $\tilde T^{\mu\nu}$ in
Eq.~(\ref{BF}) obtained from $\Theta^{\mu\nu}$ in the ideal spin
hydrodynamics in the previous section with $S^{\alpha\mu\nu}=u^\alpha
S^{\mu\nu}=\chi u^\alpha \omega^{\mu\nu}$, we work out the Landau's
condition for the local energy density and the fluid velocity, $\tilde
T^{\mu\nu}\tilde u_\nu=-\tilde \energy \tilde u^\mu$, to obtain
$\tilde\energy$ and $\tilde u^\mu$ as
$
\tilde{\energy} = \energy - \Delta\energy$,
$\tilde{u}^\mu = u^\mu -\Delta u^\mu
$
with 
\begin{equation}\label{eq:DeDu}
\Delta\energy = 2\chi \omega_{\mu}\omega^{\mu}, \quad \Delta u^\mu = -
{1\over 2\beta w}\Delta^\mu_\alpha\partial_\lambda(\beta\chi
\omega^{\alpha\lambda}).
\end{equation}
In addition, we allow a redefinition of pressure $\tilde p=p-\Delta p$ with $\Delta p=2a_0\omega_\mu\omega^\mu$, where $a_0$ is a free thermodynamic function.
In terms of these variables, the energy-momentum tensor in conventional hydrodynamics reads
\be  \label{Belinfante-RosenfeldTensor}
\tilde{T}^{\mu\nu} = \tilde{\energy} \tilde{u}^\mu\tilde{u}^\nu + \tilde p\tilde\Delta^{\mu\nu} + \tilde\tau^{\mu\nu},
\ee
where $\tilde\tau^{\mu\nu}$ denotes certain second-order transport terms 
\be
\tilde\tau^{\mu\nu} = {1\over 2}\chi\left((\theta^\mu_{\,\,\,\alpha} + \omega^\mu_{\,\,\,\alpha})\omega^{\alpha\nu} + (\mu \leftrightarrow \nu) \right)+2a_0\Delta^{\mu\nu} \omega_\lambda\omega^\lambda.\label{taumunu}
\ee
Similarly, the charge and the entropy currents in the conventional
hydrodynamics are given by
\begin{align}
\tilde j^\mu&=\tilde n\tilde u^\mu-{n\over 2\beta w}\Delta^{\mu}_\lambda \partial_\nu (\beta \chi \omega^{\lambda\nu})-{1\over 2}\Delta^{\mu}_\lambda \partial_\nu (a\omega^{\lambda\nu}), \label{TranformedChargeCurrent}\\
\tilde s^\mu&=\tilde s\tilde u^\mu-{s\Delta^{\mu}_\lambda \partial_\nu (\beta \chi \omega^{\lambda\nu}) \over 2\beta w} + {n\chi\omega^{\mu\nu}\partial_\nu\alpha \over 2w}-{\Delta^{\mu}_\lambda \partial_\nu (b\omega^{\lambda\nu})\over 2},\nonumber\\ \label{TransformedEntropyCurrent}
\end{align}
with other second-order transport terms. A similar observation was made in Ref.\cite{Fukushima:2020ucl}. It should be emphasized that the ideal limit of spin hydrodynamics with $\partial_\mu s^\mu=0$ that we start with guarantees that the conventional hydrodynamics with the above second order transport terms is also ideal, i.e. $\partial_\mu \tilde s^\mu=0$.

However, to make the conventional hydrodynamics truly conventional,
the thermodynamics relation of spin hydrodynamics in
Eq.~(\ref{ThermodynamicRelationOfSpinHydro}) should transform into
conventional thermodynamic relations,
\be  \label{conventionalThermRelation}
d\tilde{s} = \tilde{\beta}d\tilde{\energy} - \tilde{\alpha}d\tilde{n}, \quad \tilde s = \tilde{\beta}(\tilde{\energy}+\tilde p) -\tilde{\alpha}\tilde{n}.
\ee
We now show that there exists unique choice of $(a,b)$ to achieve this
equivalence, with $(a,b)$ expressed in terms of $(\chi,a_0)$
without any integrations.

We start from the entropy density $s$ in the spin
hydrodynamics as a function of density variables,
$s(\energy,n,\sigma)$, where $S^\mu\equiv
\epsilon^{\mu\nu\alpha\beta} u_\nu S_{\alpha\beta}/2$ and $\sigma\equiv S_\mu S^\mu/2$.  The
first law of thermodynamics in
Eq.~(\ref{ThermodynamicRelationOfSpinHydro}), $ds=\beta d\energy-\alpha
dn-\beta\gamma_{\mu\nu} dS^{\mu\nu}/2=\beta d\energy-\alpha
dn-\beta\gamma_{\mu} dS^{\mu}$, then gives us $\beta\equiv({\partial
s/\partial \energy})_{n,\sigma}$, $\alpha\equiv-({\partial
s/\partial n})_{\energy,\sigma}$ and $\beta\gamma_\mu\equiv
-({\partial s/\partial \sigma})_{\energy,n}S^\mu$. In local
equilibrium, $\gamma_\mu=\omega_\mu$, and the spin susceptibility is
identified as $\chi\equiv-\beta({\partial s/\partial
  \sigma})_{\energy,n}^{-1}$ from $S^\mu=\chi\omega^\mu$.

To find the
first law of thermodynamics in the conventional hydrodynamics, we
express $\tilde s$ in terms of the variables in the conventional
hydrodynamics as
\be
\tilde s(\tilde\energy,\tilde n,\omega_\mu\omega^\mu)
=s(\tilde\energy+\Delta\energy, \tilde n+\Delta n,\sigma)-\Delta s
\ee
where $\sigma={\chi^2}\omega_\mu\omega^\mu/2$, with the same
function $s$ and $(\Delta\energy,\Delta
n,\Delta s)$ given by Eqs.~(\ref{eq:DnDs}) and~(\ref{eq:DeDu}).

It is now straightforward to find the first law of thermodynamics
\be
d\tilde s=\tilde\beta d\tilde\energy-\tilde\alpha d\tilde n+A\omega_\mu d\omega^\mu
\ee
with 
\bear
\tilde\beta&=&\beta+\left(\beta\chi_\energy+b_\energy+\alpha a_\energy\right)\omega_\mu\omega^\mu, \\
\tilde\alpha&=&\alpha-\left(\beta\chi_n+b_n+\alpha a_n\right)\omega_\mu\omega^\mu, \\
A&=& 3\beta\chi+2\alpha a+2b,
\eear
where $f_{n}\equiv(\partial f / \partial n)_\energy$ and $f_{\energy}\equiv(\partial f /\partial\energy)_n$. From $s=\beta(\energy+p)-\alpha n-\beta\chi \omega_\mu \omega^\mu$ in (\ref{ThermodynamicRelationOfSpinHydro}), we also find straightforwardly
\be
\tilde s=\tilde\beta(\tilde\energy+\tilde p)-\tilde\alpha\tilde n+B\omega_\mu\omega^\mu,
\ee
with
\bear
B&=&\beta\chi+\alpha a+b-w\left(\beta\chi_\energy+b_\energy+\alpha a_\energy\right)\nonumber\\&&-n\left(\beta\chi_n+b_n+\alpha a_n\right)+2\beta a_0,
\eear
The conventional thermodynamics relations
in Eq.~(\ref{conventionalThermRelation}) are obtained by imposing the
conditions $A=B=0$. It is easy to see that these conditions determine
$(a,b)$  in terms of $(\chi,a_0)$ without any integrations, and we skip their explicit expressions.

%\be
%a={\beta\over 2\alpha_p c_s^2}\left({\chi\over w}-\beta_\energy\left({3\chi\over\beta}+\left({\partial\chi\over\partial\beta}\right)_\alpha\right)-{4 a_0\over w}\right)-{\beta\over 2} \left({\partial\chi\over\partial\alpha}\right)_\beta
%\ee
With $(a,b)$ given in terms of $(\chi,a_0)$, we see that all
second-order transport coefficients in the energy-momentum tensor,
Eq.~(\ref{taumunu}), in the charge current,
Eq.~(\ref{TranformedChargeCurrent}) and in the entropy current,
 Eq.~(\ref{TransformedEntropyCurrent}), can be expressed in terms of two free
thermodynamic functions $(\chi,a_0)$.  With the identification of
$a_1=\chi$, one can non-trivially check that these second-order
transport coefficients agree precisely with those we find in the
non-dissipative second-order hydrodynamics in the previous section
once they are also expressed in terms of $(a_1,a_0)$.
The conditions $A=0$ and $B=0$ correspond to the constraint
$C^{(5)}=0$ and a linear combination of $C^{(i)}=0$, respectively.
In the special case of conformal system,
condition $B=0$ follows from $A=0$ and conformality. This
completes the proof that the ideal spin hydrodynamics is equivalent to
the non-dissipative second-order hydrodynamics by pseudo-gauge
transformation.

\sect{Conclusion and discussion} In this \letter, we introduce a novel
power counting scheme for gradients of hydrodynamic variables and
discover nontrivial constraints on certain non-dissipative second-order
transport coefficients imposed by the second law of thermodynamics.
We also show that the spin hydrodynamics and the conventional
hydrodynamics with these second-order transport coefficients are two
equivalent descriptions of the same system related by pseudo-gauge
transformation. In a more concrete form, one can express the
hydrodynamic variables in one description in terms of those in the
other description.

Furthermore, one can construct infinitely many equivalent spin
hydrodynamics descriptions for the same system by performing
pseudo-gauge transformations using an arbitrary fraction of the spin
tensor, i.e., with $\Sigma^{\alpha\mu\nu}=tS^{\alpha\mu\nu}$, where $t\neq
1$. This transformation changes the spin susceptibility
$\chi\to (1-t)\chi\equiv \chi(t)$ in thermodynamic relations, while
$a_1(t)+\chi(t)$ remains invariant. The other second-order transport
coefficients are related to $a_1(t)$ by Eqs.(\ref{bia1}).  The
conventional hydrodynamics is a special choice in this infinite family
corresponding to $t=1$. In general, the vorticity driven thermal Hall
effect is given by Eq.(\ref{heatCurrent}) with
$\chi\to a_1(t)+\chi(t)$.

What is the meaning of spin densities in different but equivalent spin
hydrodynamics descriptions? Our results naturally suggest that the
answer to this question cannot be found within hydrodynamics
itself. For example, different species of microscopic constituents
could carry their own spins, and it is a matter of choice what to
include in the hydrodynamic description. Different choices describe
the same system, while the non-dissipative second-order transport
coefficients corresponding to each choice are related in the specific
way we described.

Finally, it would be interesting and important for practical applications to
study stability and causality issues in the class of ideal
hydrodynamic theories we considered, which would necessitate
including dissipative terms.

\acknowledgments
We thank Masaru Hongo, Xu-Guang Huang and Enrico Speranza for
discussions and P. Kovtun for bringing Refs.\cite{Bhattacharyya:2012nq,Jensen:2012jh,Banerjee:2012iz,Bhattacharyya:2014bha} to our attention.
This work is supported by the U.S. Department of Energy, Office of
Science, Office of Nuclear Physics, Grant No. DE-FG0201ER41195, and
within the framework of the Beam Energy Scan Theory (BEST) Topical
Collaboration.

\bibliographystyle{utphys}
\bibliography{SpinHydro}

\end{document}